\documentclass[twocolumn]{aastex63}
\usepackage{epstopdf}
\usepackage{amsmath}
\usepackage{amssymb}

\usepackage{color}

\shorttitle{Morphological Transformation and Star Formation Quenching}
\shortauthors{Liu et al.}
\submitjournal{ApJ}

\begin{document}

\title{Morphological Transformation and Star Formation Quenching of Massive Galaxies at $0.5 \leq z \leq 2.5$ in 3D-HST/CANDELS}
\correspondingauthor{Qirong Yuan}
\email{yuanqirong@njnu.edu.cn}

\author{Shuang Liu}
\affil{\rm Department of Physics and Institute of Theoretical Physics, Nanjing Normal University, Nanjing 210023, China}

\author{Yizhou Gu}
\affil{\rm School of Physics and Astronomy, Shanghai Jiao Tong University, 800 Dongchuan Road, Minhang, Shanghai 200240, China}

\author{Qirong Yuan}
\affil{\rm Department of Physics and Institute of Theoretical Physics, Nanjing Normal University, Nanjing 210023, China}

\author{Shiying Lu}
\affil{\rm School of Astronomy and Space Science, Nanjing University, Nanjing 210093, People's Republic of China}

\author{Min Bao}
\affil{\rm Department of Physics and Institute of Theoretical Physics, Nanjing Normal University, Nanjing 210023, China}
\affil{\rm School of Astronomy and Space Science, Nanjing University, Nanjing 210093, People's Republic of China}

\author{Guanwen Fang}
\affil{\rm School of Mathematics and Physics, Anqing Normal University, Anqing 246011, China}

\author{Lulu Fan}
\affil{\rm CAS Key Laboratory for Research in Galaxies and Cosmology, Department of Astronomy, University of Science and Technology of China, Hefei 230026, China}
\affil{\rm School of Astronomy and Space Sciences, University of Science and Technology of China, Hefei, Anhui 230026, People's Republic of China}

\begin{abstract}
To figure out the effect of stellar mass and local environment on morphological transformation and star formation quenching in galaxies, we use the massive ($M_* \geq 10^{10} M_{\odot}$) galaxies at $0.5 \leq z \leq 2.5$  in five fields of 3D-HST/CANDELS. Based on the {\it UVJ} diagnosis and the possibility of possessing spheroid, our sample of massive galaxies are classified into four populations: quiescent early-type galaxies (qEs), quiescent late-type galaxies (qLs), star-forming early-type galaxies (sEs), and star-forming late-type galaxies (sLs). It is found that the quiescent fraction is significantly elevated at the high ends of mass and local environmental overdensity, which suggests a clear dependence of quenching on both mass and local environment. Over cosmic time, the mass dependence of galaxy quiescence decreases while the local environment dependence increases. The early-type fraction is found to be larger only at high-mass end,  indicating a evident mass dependence of morphological transformation. This mass dependence becomes more significant at lower redshifts. Among the four populations, the fraction of active galactic nucleus (AGN) in the qLs peaks at $2<z \leq 2.5$, and rapidly declines with cosmic time. The sEs are found to have higher AGN fractions of $20-30\%$ at $0.5\leq z<2$ . The redshift evolution of AGN fractions in the qLs and sEs suggests that the AGN feedback could have played important roles in the formation of the qLs and sEs.
%which might be responsible for truncating the star formation and maintaining the quiescence for the qLs at $z>2$, and triggering the starbursts for the sEs at $z<2$.

\end{abstract}

\section{Introduction} \label{sec1:intro}

Red early-type and blue late-type galaxies are two distinct populations in the color-magnitude diagrams from the local universe to $z \sim 3$, which are connected with stellar mass and kinematics \citep{York+2000,Strateva+2001,Stoughton+2002,Wyder+2007,Whitaker+2011,Tomczak+2014,Mortlock+2015,Straatman+2016,Davidzon+2017}.
In the local universe, early-type galaxies are usually found to be red ellipticals or bulge-dominated lenticulars with little star formation, while typical late-type galaxies exhibit disk-dominated appearance with spiral arms and intensive star formation \citep{Baldry+2004}. It is evidenced that the bimodal distributions are still observed at $z > 1$. The abundance of blue late-type galaxies tends to decrease over cosmic time, while red early-type galaxies become abundant since $z \sim 1$ \citep{Strateva+2001,Bell+2004,van_Dokkum+2006}.

Stellar mass and environment are two important factors to influence the evolution of galaxies. Massive galaxies possess higher specific star formation rates (sSFRs) at earlier times, and have a higher probability to be quenched than less massive galaxies, which is named as the ``downsizing'' scenario \citep{Bell+2003,Kauffmann+2003,Noeske+2007,Gu+2018}. Different environmental properties can measure the level of galaxy interaction and dependence of AGN \citep{Larson+1980, Moore+1999, Moran+2007, Li+2019}. Red early-type galaxies are also more inclined to locate at denser environment toward lower redshifts which suggests that environmental density is effective to star formation quenching as time proceeds \citep{Bundy+2006,Cooper+2008,Tasca+2009,Peng+2010,Gu+2018}.

The general picture has been established that the transformation from disk-dominated star-forming galaxies(SFGs) to the quiescent early types prefers to happen in the galaxies with higher stellar mass and in denser environment,  coupling with the star formation quenching \citep{Bundy+2010,Gu+2018}. Morphological transformation and star formation quenching are usually thought to have a degeneracy, which means that the shutdown of star formation might be linked with the buildup of bulge structure.  Merger and disk instability is able to boost the morphological change via rapid bulge growth, and to trigger the feedback from active galactic nucleus (AGN) which heats or drives out the cold gas \citep{Croton+2006,Faber+2007, Hopkins+2008, Schawinski+2014, Brennan+2017}. On one hand, some violent activities in a denser environment, such as galaxy-galaxy merger, tidal stripping, and galaxy harassment can efficiently transform the morphologies of galaxies and quench their star formation \citep{TT+1972,WR+1978,Moore+1996,Lake+1998,Hopkins+2008,KH+2013}. On the other hand, some mild processes such as strangulation and AGN heating can also cease the star formation in galaxies by cutting off the accretion of cold gas or suppressing gas from further cooling without destroying their disks \citep{Larson+1980,Fabian+1994,Balogh+2000,Weinmann+2006,Bekki+2009}.

The emergence of red spiral (i.e., passive late-type) and blue elliptical (i.e., star-forming early-type) galaxies breaks the degeneracy between star formation quenching and morphological transformation, which provides the possibility to figure out the two processes individually. The blue ellipticals are spheroidal-dominated galaxies with active star formation. Their formation might be attributed to the major merger on the basis of their high asymmetry indices \citep{Liu+2019}. In addition, the rejuvenation of red elliptical galaxies may also trigger star formation again at low mass by galaxy mergers \citep{Kannappan+2009,Kim+2018}.
Red spiral galaxies possess little star formation, but their disks still remain. Some recent efforts have been devoted to evaluate their physical properties and origins \citep{Bundy+2010,Masters+2010}. The abundance of red spiral galaxies suggests that they are not confined to a single process among environmental effect, disk regrowth and  internal instability at $z \sim 1-2$ \citep{Bundy+2010}. \cite{Fraser-McKelvie+2018} proposed that massive and less massive red spiral galaxies may have experienced different processes: less massive galaxies are inclined to locate in clusters where ram-pressure stripping and strangulation are responsible for the star formation quenching; although massive red spiral galaxies  possess higher bar fractions than star-forming counterparts, their quenching mechanism is less clear. \cite{Hao+2019} accessed their two-dimensional spectroscopy of blue and red spirals at $z \sim 1$. Similar to red ellipticals, red spiral galaxies are found to have older stellar populations and higher stellar metallicities than blue spirals. These red spirals are verified to be remnants of gas-rich mergers.

To understand the connection between morphology and star formation, a number of studies have been carried out in this regard. \cite{Liu+2019} considered the quenching of star formation and the transformation of morphology independently by classifying the galaxies at $0.01 < z < 0.12$ into four types: quenched early-type galaxies (qEs), quenched late-type galaxies (qLs), star-forming early-type galaxies (sEs) and star-forming late-type galaxies (sLs).
They found that morphological transformation is mainly regulated by stellar mass among various properties including stellar mass, halo mass, halo radius, and environment. The quenching of star formation is found to be mainly driven by stellar mass for more massive galaxies, but by halo mass for the lower-mass galaxies.
Using the IllustrisTNG simulation, \cite{Tacchella+2019} found that galaxy morphologies are basically organized during the phase of active star formation, and the bulge formation of intermediate-mass galaxies is mainly driven by mergers.

In order to address the morphological transformation and star formation quenching over a long period of cosmic time, we aim to focus on the massive ($M_* \geq 10^{10} M_{\odot}$) galaxies at $0.5 \leq z \leq 2.5$ in five 3D-HDT/CANDELS fields. Following \cite{Liu+2019}, we also classify the galaxies into four populations (i.e., qEs, qLs, sEs, and sLs), and then analyze how their abundances vary with stellar mass and environment. Since AGN feedback might have played a role on morphological transformation and star formation quenching, AGN fraction is also taken into consideration. %Besides this, we estimate the timescales of quenching and morphological transformation. 
The implications are also discussed regarding to the formation of sEs and qLs.

This paper is structured as follows. We describe the data set of 3D-HST/CANDELS program and our sample construction in Section \ref{sec2:sample}. The distributions of stellar mass and environmental overdensity for these four galaxy populations are shown in Section \ref{sec3:mass and environment distributions}. The dependence of star formation quenching on mass and environment is presented in Section \ref{sec4:star formation quenching}, while their effects of mass and environment on morphological transformation are given in Section \ref{sec5:morphology transformation}. Possible roles of AGN feedback on the formation of qLs and sEs are discussed in Section \ref{sec6:AGN fractions}. In Section \ref{sec7:discussion}, we try to explain our findings from a viewing angle of the timescales of morphological transformation and star formation quenching. Finally, a summary is given in Section \ref{sec8:summary}. We assume the following cosmological parameters throughout the paper:  $H_{0} = 70{\rm\ km}{\rm\ s^{-1}\ Mpc^{-1}}$, $\Omega_{\rm m} = 0.30$, $\Omega_{\Lambda} = 0.70$. All magnitudes given in this paper are in the AB system.

\section{Data and Sample Selection} \label{sec2:sample}

Covering $900~ {\rm arcmin}^{2}$ area in five fields (namely, AEGIS, COSMOS, GOODS-N, GOODS-S and UDS), the 3D-HST and CANDELS  programs have provided a wealth of multi-wavelength database acquired by the WFC3 and ACS spectroscopy and photometry, which makes it possible to build the spectral energy distributions (SEDs) from ultraviolet to infrared bands \citep{Grogin+2011,Koekemoer+2011, Skelton+2014}.

Based on previous ground-based spectroscopic surveys, the available spectroscopic redshifts ($z_{\rm spec}$) in five fields are collected, while photometric redshifts ($z_{\rm phot}$) are derived by \cite{Skelton+2014} using the $0.3 - 8.0~ \mu m$ SEDs in EAZY code \citep{Brammer+2008}. Besides, \cite{Momcheva+2016} explored the grism observations and derived the ``best'' redshifts ($z_{\rm best}$) through merging grism spectrum redshifts ($z_{\rm grism}$) fits with the photometric and spectroscopic redshifts given by \cite{Skelton+2014}. The normalized median absolute deviations of $z_{\rm phot}$, defined by $\sigma_{\rm NMAD} = 1.48 \times {\rm median}[|(z_{\rm phot}-z_{\rm spec})-{\rm median}(z_{\rm phot}-z_{\rm spec})|/(1+z_{\rm spec})]$, are 0.022, 0.007, 0.026, 0.010, and 0.023 for the AEGIS, COSMOS, GOODS-N, GOODS-S, and UDS fields, respectively \citep{Skelton+2014}. While the typical redshifts error for $z_{\rm grism}$ is $\sigma_{z} \approx 0.003 \times (1 + z)$, indicating that $z_{\rm grism}$ is apparently of higher accuracy than $z_{\rm phot}$.
In this paper, the ``best'' redshifts ($z_{\rm best}$) are adopted, which means that we prefer to take the $z_{\rm spec}$ and $z_{\rm grism}$, otherwise the $z_{\rm phot}$ will be used instead.

Applying the well-constrained redshifts (i.e., $z_{\rm best}$), \cite{Momcheva+2016} have derived the stellar population parameters by using FAST code \citep{Kriek+2009}, assuming the exponentially declining star formation history with the e-folding time scale $\tau = 0.1 - 10$ Gyr and dust attenuation $A_{V} = 0 - 4$ in the \cite{Calzetti+2000} reddening law. They adopted the \cite{BC+2003} stellar population synthesis model library with a \cite{Chabrier+2003} initial mass function and solar metallicity. The derived stellar population parameters include the stellar mass ($M_{\ast}$), the dust attenuation ($A_{V}$) and the stellar ages.

Based on the released database, we only choose the galaxies with a flag of ``{\tt use{\_}phot} = 1'', which ensures that the target source is a reliable exposure with good signal-to-noise ratio, and not surrounded by bright stars (see \citealt{Skelton+2014} for more information).

\subsection{\rm Star-forming vs. Quiescent}\label{sec2.1:star forming vs. quiescent}
In previous studies, the state of star formation activity has usually been diagnosed by the $UVJ$ diagram (i.e., rest-frame $U - V$ vs. rest-frame $V - J$). The $UVJ$ diagram is now widely employed to distinguish SFGs and quiescent galaxies(QGs) at high redshifts \citep{Wuyts+2007,Williams+2009,Straatman+2016,Fang+2018}. The rest-frame $U - V $ color can separate blue and red colors on account of different levels of star formation. Yet blue populations will also appear redder due to the presence of a large amount of dust and gas at high redshifts. Therefore, dust attenuation should be considered by using the dust-absorbed infrared $J$-band to relieve this situation. In this work, we adopt the following criteria for separating the QGs from SFGs \citep{Williams+2009}:
\begin{eqnarray}
&&(U - V) > 1.3,(V - J) < 1.6,    \\
&&(U - V) > 0.88 \times (V - J) +0.49\ (0.5 < z < 1.0),   \\
&&(U - V) > 0.88 \times (V - J) +0.59\ (1.0 < z < 2.5),
\end{eqnarray}
where QGs reside in the wedged region while SFGs are scattered in the remaining area of the $UVJ$ diagram. %Therefore, the number of quiescent galaxies in our total galaxies is 2931, and the rest are 6619 star-forming galaxies.

%Based on the released database, we firstly construct a large parent sample of 9550 massive ($ M_* \geq 10^{10}M_\odot$) galaxies at $0.5 \leq z \leq 2.5$. The mass threshold ($ M_* \geq 10^{10}M_\odot$) guarantees a high completeness of $\sim$90\% out to $z \sim 3$, even for the quiescent galaxies(QGs) which are more difficult to be detected \citep{Grogin+2011, Wuyts+2011, Newman+2012, Barro+2013, Pandya+2017}.

In a magnitude-limited sample, the stellar mass completeness depends on both the redshift and mass-to-light ratio ($M/L$). With higher $M/L$, QGs are more difficult to be detected \citep{Grogin+2011, Wuyts+2011, Newman+2012, Barro+2013, Pandya+2017}. Higher completeness of QGs guarantees the completeness of SFGs. In \cite{Chartab+2020}, following the methodology of \cite{Pozzetti+2010}, it had proved that the mass threshold ($M_* \geq 10^{10}M_\odot$) can ensure a 95\% completeness with magnitude limit $H_{\rm lim} < 26$ for QGs. But, the error of $z_{\rm phot}$ will increase from 0.002 to 0.0046 at $H_{\rm lim} = 25-26$ \citep{Bezanson+2016}. To ensure a higher completeness and accuracy of $z_{\rm phot}$, we construct a magnitude-limit sample ($H_{\rm lim} < 25$) of 9550 massive galaxies with $ M_* \geq 10^{10}M_\odot$ at $0.5 \leq z \leq 2.5$. As illustrated in Figure~1 of Gu et al. (2021, resubmitted), for magnitude-limited sample of $H_{\rm lim}<25$, this mass threshold can guarantee a 90\% completeness of QGs at $0.5<z<2.0$. However, at $2.0<z<2.5$, we remind that this mass threshold might only represent $\sim$80\% completeness.

Due to small areas of the coverage for five independent fields of 3D-HST/CANDELS survey, the cosmic variance might be a significant source of uncertainty \citep{Meneux+2009}. In the halo occupation models at different redshifts, the relationship between stellar mass and dark matter halo(subhalo) mass is empirically established, and the galaxy bias can be calculated by using a dissipationless N-body simulation \citep{Moster+2010,Moster+2011}. Applying the recipe described by \cite{Moster+2011}, we estimate the cosmic variance of massive ($\log M_*/M_{\odot} \geq 10$) galaxies in each of the four redshift intervals (i.e., $0.5<z<1.0$, $1.0<z<1.5$, $1.5<z<2.0$, $2.0<z<2.5$), and their typical uncertainties are 16\%, 16\%, 18\%, and 21\% respectively. In \cite{Huertas-Company+2016}, it is also evidenced that the cosmic variance tends to become larger at higher redshifts and stellar mass bins. In the following discussion, we need to keep in mind however that our results at high redshifts might attribute to higher uncertainty.

%Making use of the galaxy bias, this uncertainty can be predicted for a given galaxy population using a halo occupation models provided by \cite{Moster+2011}.

%As roughly estimated by \cite{Bundy+2006}, while the absolute comparisons between different redshift intervals are largest affected by this cosmic variance uncertainty, the comparisons of relative or fraction abundance for a given population are less susceptible to it. }}

\subsection{\rm Early-type vs. Late-type} \label{sec2.2:Early type vs. Late type}

\begin{figure*}
  \centering
  \includegraphics[width=18cm]{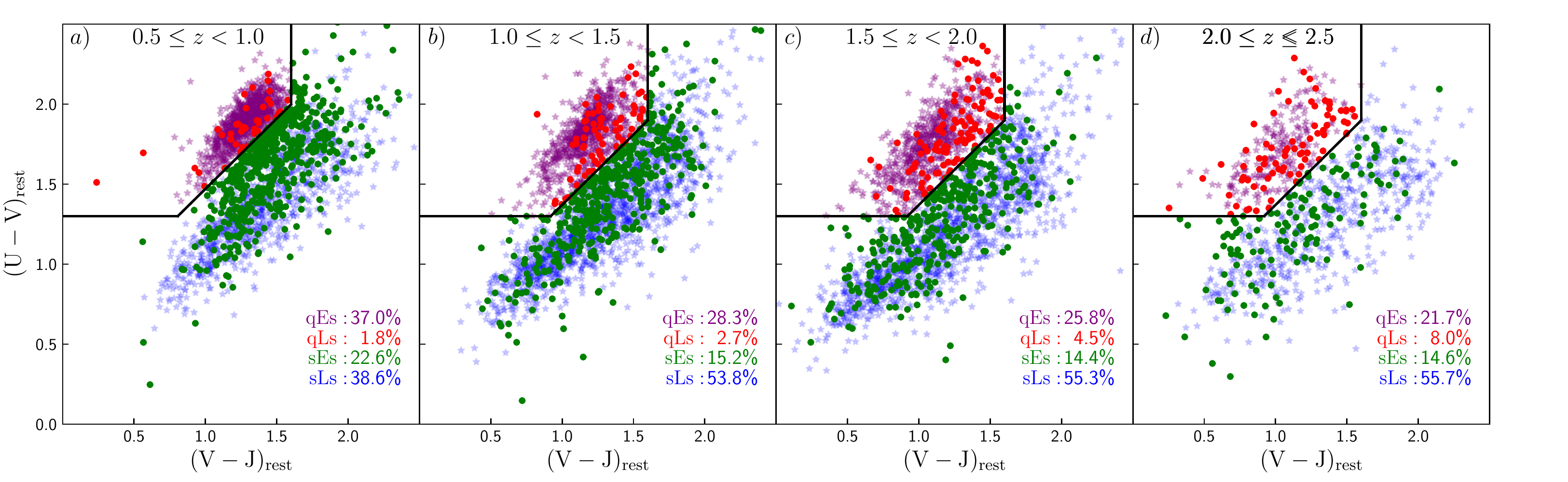}\\
  \caption{The distributions of qEs, qLs, sEs and sLs in the $UVJ$ diagram at different $z$-bins. Purple and blue colors denote the qEs and sLs respectively, while the qLs and sEs are represented in red and green dots. The fractions for four subsamples out of total galaxies are shown in right-bottom corner of each panel.}\label{fig1_UVJ}
\end{figure*}

\begin{table*}
  \caption{The numbers of qEs, qLs, sEs and sLs at different $z$-bins}\label{table1:numbers of four types}
  \begin{tabular}{c|cccc|cc|cc}
  \hline
  \hline
  Redshifts      &qEs    &qLs     &sEs      &sLs     &early-type   &late-type   &quiescent   &star-forming  \\
  \hline
  $0.5 \leq z <1.0$        &896    &43      &547      &935       &1443      &978     &939     &1482     \\
  $1.0 \leq z <1.5$        &781    &74      &418      &1485      &1199      &1559    &855     &1903    \\
  $1.5 \leq z <2.0$        &625    &109     &348      &1336      &973       &1445    &734     &1684      \\
  $2.0 \leq z \leq 2.5$    &265    &96      &319      &1113      &584       &1209    &361     &1432   \\
  \hline
  \end{tabular}
\end{table*}

For magnitude-limited samples of visual-like H-bands ($H_{\rm F160W} < 24.5$) in the CANDELS five fields, released in Rainbow database \footnote{\url{http://rainbowx.fis.ucm.es/Rainbow\_navigator\_public/}}, \cite{Huertas-Company+2015} have utilized the Convolutional Neural Networks (ConvNets) machine-learning algorithm to investigate the morphologies of $\sim 50,000$ galaxies with median redshifts of $\langle z \rangle \sim 1.25$. The algorithm is trained by the visual classification publicly available in GOODS-S field, and then applied to the other four fields to make a successful classification. For each galaxy, the possibilities of holding a spheroid or a disk, being irregular, point source or unclassified ($f_{\rm sph}$, $f_{\rm disk}$, $f_{\rm irr}$, $f_{\rm ps}$ or $f_{\rm unc}$) are estimated. The mis-classification rate is less than 1\%, and this is a major step forward compared with the other Concentration-Asymmetry-Smoothness (CAS)-based methods which yields 20\%-30\% contamination at high redshifts \citep{Huertas-Company+2014}. They also have provided possible thresholds of different frequencies to classify galaxy morphology and calibrate with visual inspection, which are:
\begin{enumerate}
\item pure bulges [SPH]:\\ $f_{\rm sph} > 2/3$,\qquad$f_{\rm disk} < 2/3$,\qquad and $f_{\rm irr} < 1/10$;
\item disk+sph [DSPH]:\\ $f_{\rm sph} > 2/3$,\qquad$f_{\rm disk} > 2/3$,\qquad and $f_{\rm irr} < 1/10$;
\item pure disk [DISK]:\\ $f_{\rm sph} < 2/3$,\qquad$f_{\rm disk} > 2/3$,\qquad and $f_{\rm irr} < 1/10$;
\item irregular disks [DIRR]:\\ $f_{\rm sph} < 2/3$,\qquad$f_{\rm disk} > 2/3$,\qquad and $f_{\rm irr} > 1/10$;
\item irregulars/mergers [IRR]:\\  $f_{\rm sph} < 2/3$,\qquad$f_{\rm disk} < 2/3$, and\qquad $f_{\rm irr} > 1/10$.
\end{enumerate}

In our sample, a vast majority of massive galaxies ($\sim 90 \%$) can be  successfully classified into above morphologies. The remaining galaxies are either with fainter magnitudes in H-band or not included into above criteria (e.g., $f_{\rm sph}>2/3$ AND $f_{\rm irr}>0.1$). For these unclassified galaxies with $H_{\rm F160W} < 24.5$, we further classify them into above five typical morphologies with eyeballing inspection, which leads to about 98\% classification rate in our sample. Then we simplify the morphological types as early-type (including the SPH and DSPH) and late-type (including the DISK, DIRR and IRRs) galaxies, which depends on whether they have a dominated spheroidal component (i.e., $f_{\rm sph} > 2/3$ ) or not.
%With eyeballing inspection, we further classify the remaining galaxies into above five morphologies by ourselves, which improves the successful classification rate to 98\% both for quiescent and star-forming galaxies.Unclassified galaxies are either with fainter magnitude in H-band or locate at the boundary of coverage.

Accounting to above criteria, the sample of massive galaxies are divided into four subsamples: (1) 2567 quiescent early-type galaxies (qEs), (2) 322 quiescent late-type galaxies (qLs), (3) 1632 star-forming early-type galaxies (sEs), and (4) 4869 star-forming late-type galaxies (sLs).  As mentioned above, only 2\% galaxies are discarded which is hard to distinguish via visual inspection due to the fainter brightness ($H_{\rm F160W} > 24.5$).
The detailed numbers of four galaxy types in different redshift ranges are listed in Table \ref{table1:numbers of four types}.
The locations in the $UVJ$ diagram for four galaxy types are presented in Figure \ref{fig1_UVJ}, and the fractions of four subsamples out of total galaxies is shown in the lower right corner of each panel. Obviously, it is distinct that both the fractions of early-type (qEs $+$ sEs) and quiescent (qEs $+$ qLs) galaxies tend to increase over cosmic time, while the fractions of late-type (qLs $+$ sLs) and star-forming (sEs $+$ sLs) galaxies decrease from high to low redshifts. On one hand, it is clear to see the popularity of two traditional bimodal galaxies (i.e., qEs and sLs), suggesting that the star formation quenching in the majority of galaxies is accompanied with the growth of their spheroidal components. On the other hand, the rarity of qLs and sEs indicates small probability to break the degeneracy between morphological transformation and star formation quenching.

\section{Mass and environment distributions}\label{sec3:mass and environment distributions}
To link the mass and environment factors to star formation and morphology of galaxy, we investigate the mass and environment distribution for four types of galaxies at first.

\subsection{\rm Stellar Mass Distribution}\label{sec3.1:stellar mass distribution}

\begin{figure*}
  \centering
  % Requires \usepackage{graphicx}
  \includegraphics[width=18cm]{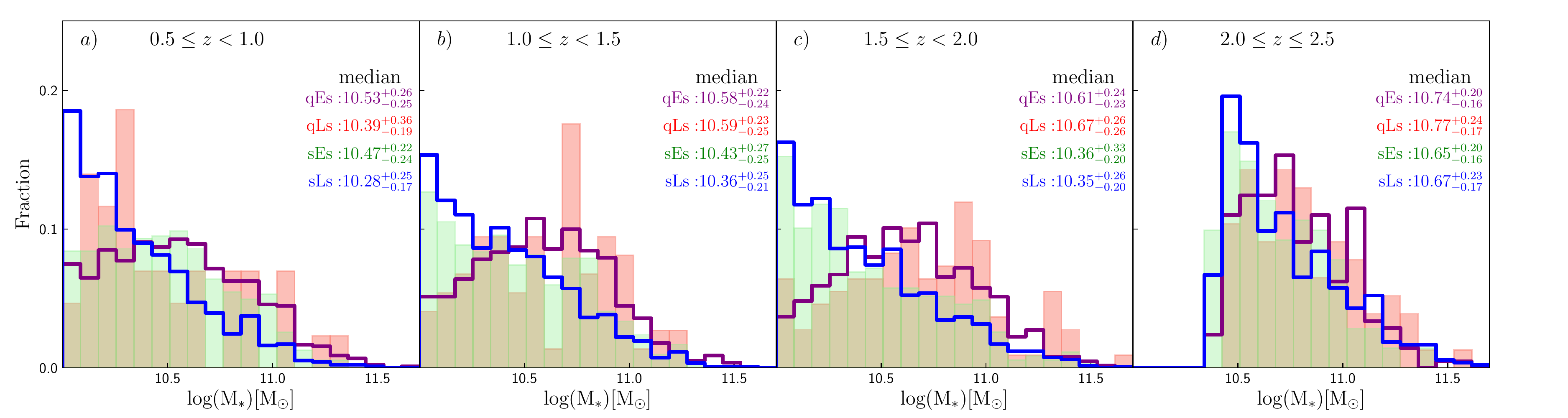}\\
  \caption{The stellar mass distributions of qEs, qLs, sEs and sLs, which are colored in purple, red, green and blue respectively. The median value and their percentile ranges from 25\% to 75\% are shown in the upper right corner of each panel.}\label{fig2_M}
\end{figure*}

\begin{deluxetable*}{ccccc}
\tablewidth{0pt}
\tablecaption{The probability of K-S test for stellar mass distributions}\label{table2:K-S test of M}
\tablehead{
 \colhead{Subsamples} & \colhead{$0.5 \leq z < 1.0$} & \colhead{$1.0 \leq z < 1.5$} & \colhead{$1.5 \leq z < 2.0$}  & \colhead{$2.0 \leq z \leq 2.5$}}
\startdata
qEs vs. sEs  &  0.017     &$1.491\times10^{-8}$    &$3.427\times10^{-16}$      &$1.195\times10^{-19}$     \\
qLs vs. sLs  &  0.138     &$5.377\times10^{-6}$    &$1.969\times10^{-10}$     &$5.184\times10^{-13}$     \\
qEs vs. qLs  &  0.058     &0.485                   &   0.147                  &   0.798                  \\
sEs vs. sLs  &$5.030\times10^{-14}$   &$0.003$    &   $0.472$     &   0.593    \\
\hline
\enddata
%\tablenotetext{}{\textbf{References.}(1) \citealt{Greiss2012}; (2)\citealt{Greiss2012}}
\end{deluxetable*}

Mass distributions of four galaxy types in each redshift bin are presented in Figure \ref{fig2_M}. Median values and their corresponding 25-75th percentile ranges are shown in the top right corner of each panel. Furthermore, to identify their distributional differences, we perform the Kolmogorov-Smirnov (K-S) test for comparing stellar mass distributions of two subsamples (see Table \ref{table2:K-S test of M}). The probability that two subsamples in a given redshift range are drawn from the same underlying distributions is referred to as the quality $P$. We adopt $P = 0.05$ as the upper limit probability to verify that the two subsamples have different distributions at $2\sigma$ deviation.

To investigate the connection between stellar mass and star formation quenching, we firstly compare the stellar mass distribution of SFGs with that of QGs at a given morphological type (i.e., qEs vs. sEs and qLs vs. sLs). According to their median values in Figure \ref{fig2_M}, QGs populate at higher stellar mass compared with their star-forming counterparts. Since the probability of star formation quenching increases with the growth of stellar mass \citep{Peng+2010,Brammer+2011,Muzzin+2013}, it is conceivable that quiescent population is more inclined to be found at high-mass end. Moreover, the significance of differences between their distributions are shown in the first two rows of Table \ref{table2:K-S test of M}. It can be found that the quality $P$ is increasing with cosmic time but still below 0.05, except for the K-S test between qLs and sLs at $0.5 \leq z < 1.0$ ($P$ = 0.138). At a given morphological type, SFGs and QGs present vastly different mass distributions at higher redshifts ($z>1$), which hints that the quenching process of galaxy star formation is likely to have been accompanied by the assembly of stellar mass. %Their differences tend to decrease with cosmic time.

To correlate stellar mass with morphological transition, the K-S tests of stellar mass distribution between early-type galaxies (ETGs) and late-type galaxies (LTGs) in the star-forming or quiescent population are performed, as shown in the last two rows of Table \ref{table2:K-S test of M}.
At $z>1.5$, no significant difference is found between the mass distributions of ETGs and LTGs in both two populations.
Although the similar mass distributions are found for qEs and qLs at $z<1.5$,  an obvious distributional difference is shown for sEs and sLs.
Hence, it indicates that the growth of spheroidal component has a closer connection with stellar mass at lower redshifts, which reinforces the notion that morphological transition is an in-situ process in the local universe \citep{Bamford+2009,Liu+2019}.

\subsection{\rm Environment Distribution} \label{sec3.2:environment distribution}

Environment is a crucial external factor for galaxy evolution \citep{Ilbert+2013,Darvish+2015,Joshi+2020}, so it is necessary to take the environmental condition around a galaxy into consideration. Traditional environment indicator is defined as $\Sigma_{N} = N/(\pi d_{N}^{2})$ by \cite{Dressler+1980}, which describes the local number density around a target within an area defined by the projected distance of the $N$-th nearest neighbor ($d_N$) within a given redshfit slice. Alternatively, a modified environment indicator has been introduced by \cite{Ivezic+2005} and \cite{CI+2008}, which is based on a Bayesian metric to incorporate the distances of all neighboring galaxies. They also evidenced that the probability distribution functions of environment by this Bayesian metric is more approach to the true distributions (see Figure 9 in \citealt{Ivezic+2005}).

Inspired by \cite{Ivezic+2005} and \cite{CI+2008}, we utilize the method of Bayesian metric to improve our measurement of environment (Gu et al., 2021, submitted). For a magnitude-limited sample of galaxies at $0.5 \leq z \leq 2.5$ with $H_{\rm F160W} < 25$, the local density is estimated by $\Sigma_{N}^{'} \propto 1/(\Sigma_{i=1}^{N}d_{i}^2)$, where $d_i$ is the projected distance to the $i$-th nearest neighbor within a redshift slice ($|\Delta z|<\sigma_z (1+z),\, \sigma_z=0.02$). A dimensionless  overdensity, $1 + \delta_{N}^{'}$, is employed to measure the relative density of environment, which described as

\begin{equation}\label{func4_dencity}
  1 + \delta_{N}^{'} = \frac{\Sigma_{N}^{'}}{\langle\Sigma_{N}^{'}\rangle_{\rm uniform}}=\frac{\Sigma_{N}^{'}}{k_{N}^{'}\Sigma_{\rm surface}},
\end{equation}
where $\langle\Sigma_{N}^{'}\rangle_{\rm uniform}$ is the Bayesian density in uniform condition. The $k_{N}^{'}$ is a correction factor of proportional relation between $\langle\Sigma_{N}^{'}\rangle_{\rm uniform}$ and surface number density $\Sigma_{\rm surface}$ within a given redshift slice (e.g., $N = 3$, $k_{3}^{'} = 0.80$). Obviously,
$1+\delta_{N}^{'}>1$ (i.e., $\log(1 + \delta_{N}^{'})>0$) implies the excess of standard level of environment density while 1 + $\delta_{N}^{'}<1$ indicates the opposite. In this paper, we prefer to apply the local density of three closest galaxies (1 + $\delta_{3}^{'}$) to indicate the nearest small-scale(local) environmental density for each targeted galaxy. It also has been confirmed that the neighboring number we adopt from 3 to 10 will not affect our main results. Moreover, following \cite{Kawinwanichakij+2017}, we test the ``edge effect'' by excluding galaxies near the survey edge ($\sim 2.95\%$). Our main results are not affected by this effect. 

\begin{figure*}
  \centering
  % Requires \usepackage{graphicx}
  \includegraphics[width=18cm]{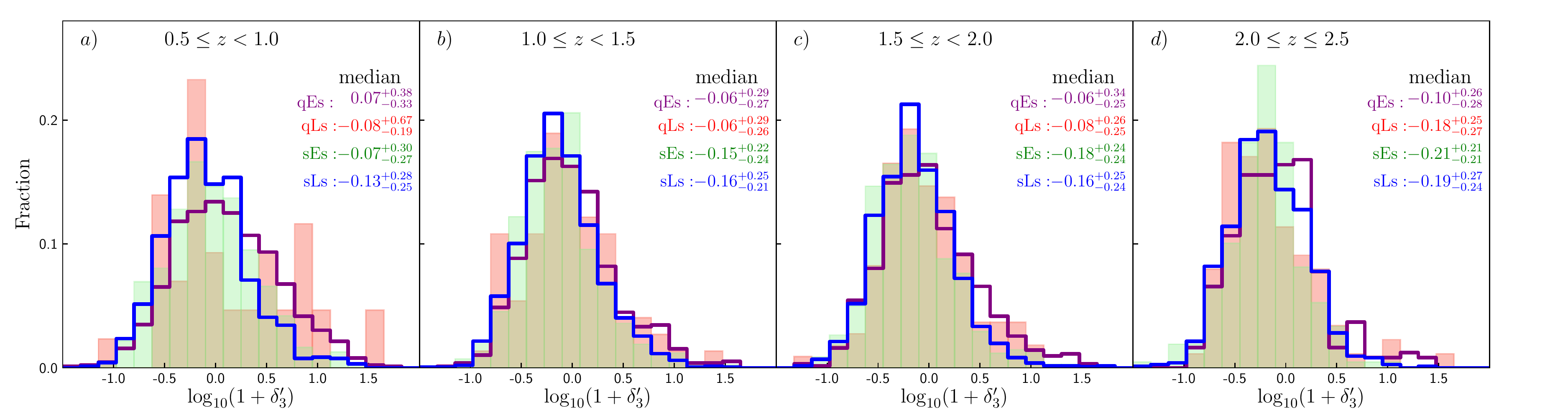}\\
  \caption{The overdensity distributions of qEs, qLs, sEs and sLs in each $z$-bin. Also given are their median values and their percentile ranges from 25\% to 75\% as Figure \ref{fig2_M}.}\label{fig3_overdensity}
\end{figure*}

\begin{deluxetable*}{ccccc}
\tablecaption{The probability of K-S test for environment distributions}\label{table3:K-S test of density}
\tablehead{
 \colhead{Subsamples} & \colhead{$0.5 \leq z < 1.0$} & \colhead{$1.0 \leq z < 1.5$} & \colhead{$1.5 \leq z < 2.0$}  & \colhead{$2.0 \leq z \leq 2.5$}}
\startdata
qEs vs. sEs   &  $2.758\times10^{-5}$      &$8.985\times10^{-4}$    &$4.437\times10^{-4}$    &     0.001     \\
qLs vs. sLs         &   0.019                    &    0.022         &    0.425               &     0.817     \\
qEs vs. qLs         &   0.429                    &    0.843         &    0.387               &     0.323     \\
sEs vs. sLs         &   0.039                    &    0.730         &    0.462               &     0.565     \\
\enddata
%\tablenotetext{}{\textbf{References.}(1) \citealt{Greiss2012}; (2)\citealt{Greiss2012}}
\end{deluxetable*}

Figure \ref{fig3_overdensity} shows the distributions of the local overdensity for four galaxy types in different redshift bins.
If we consider the quenching process of star formation for a given morphological type (i.e., early- or late-type), compared with their median values (say qEs vs. sEs, or qLs vs. sLs), it can be found that QGs locate at denser environment than their star-forming counterparts. This result confirms the color-density relation in literatures
 (\citealt{Lewis+2002,Kauffmann+2004,Rojas+2005,Weinmann+2006,Bamford+2008,Liu+2015,Moorman+2016}), suggesting that the dense environment may have played an important role in the quenching process via various modes, such as ram-pressure stripping \citep{GG+1972}, starvation \citep{Larson+1980}, and tidal interaction \citep{Merritt+1983}.
The corresponding probabilities of K-S test for environmental distributions are tabulated in the first two rows of Table \ref{table3:K-S test of density}. It can be seen that qEs and sEs present entirely different distributions of environment at $0.5 \leq z \leq 2.5$. The qLs and sLs at high redshifts ($1.5<z<2.5$) seem to reside in similar environment. However, significant environmental difference is displayed between the qLs and sLs at $0.5<z<1.5$. Therefore, it may imply that the star formation quenching is sensitive to local overdensity for both early- and late-type populations at $z<1.5$, which is in agreement with the results in \cite{Gu+2018} that the environment distribution of red galaxies differs from those of green and blue populations since $z \sim 1.5$.

Next, we analyze the environmental role on galaxy morphological transition for SFGs (including sEs and sLs) and QGs (including qEs and qLs). According to the probabilities of K-S test shown in the last two rows of Table \ref{table3:K-S test of density}, there is no significant difference between the ETGs and LTGs at $0.5\leq z \leq 2.5$, except for the star-forming populations (i.e., sEs vs. sLs) at $0.5<z<1.0$ ($P$ = 0.039). This implies that environment may not be the important factor for morphological transformation at $z>1$. For the SFGs at low redshifts ($0.5<z<1$),  sEs are likely to reside in denser environment when compared with sLs.

\section{Star formation quenching}\label{sec4:star formation quenching}

\begin{figure*}
  \centering
  % Requires \usepackage{graphicx}
  \includegraphics[width=18cm]{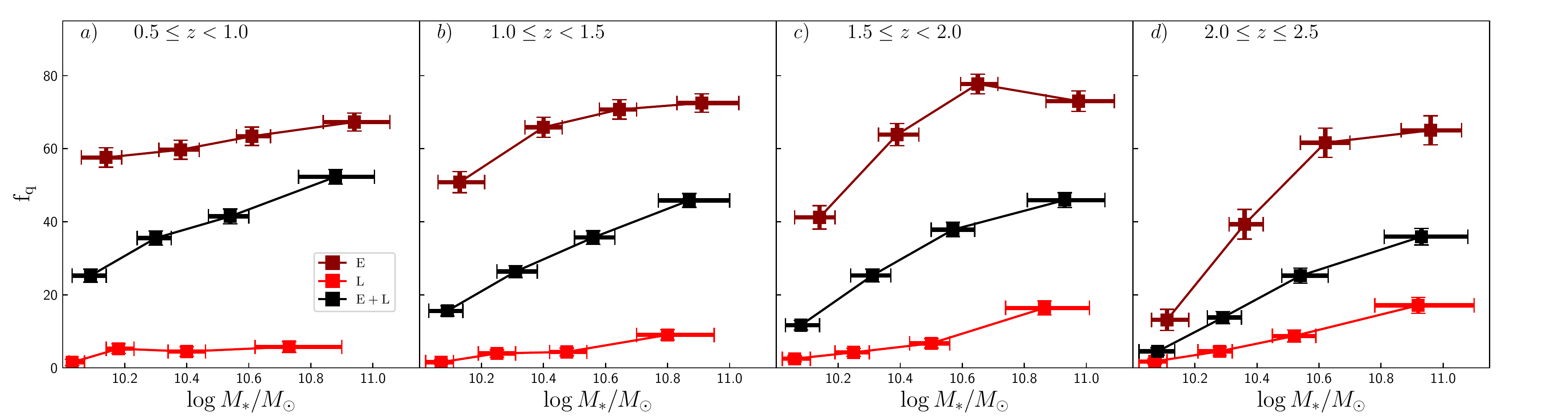}\\
  \caption{Fractions of quiescent population out of ETGs ($ f_{\rm q}\rm (E) = \frac{N_{\rm qE}}{N_{\rm E}}$), LTGs ($f_{\rm q}\rm (L) = \frac{N_{\rm qL}}{N_{\rm L}}$) and both types of galaxies ($f_{\rm q} \rm (E+L) = \frac{N_{\rm qE}+N_{\rm qL}}{N_{\rm E}+N_{\rm L}}$) as a function of stellar mass in four redshift bins with $\Delta z = 0.5$. The $f_{\rm E}$,$f_{\rm L}$,$f_{\rm (E+L)}$ are denoted by darkred, red and black colors. At each panel, stellar mass is divided into four bins with roughly equal galaxy numbers. Within each stellar mass bin, the x-axis error represents the 25th to 75th percentiles. Considering a binomial distribution, the fraction uncertainty within is calculated as $\sigma_f = [f(1 - f)/N_{\rm gal}]^{1/2}$, where the $f$ represents $f_{\rm q}(\rm E)$, $f_{\rm q}{\rm(L)}$ and $f_{\rm q} \rm(E+L)$ while the $N_{\rm gal}$ represents $N_{\rm E}$, $N_{\rm L}$, and ($N_{\rm E}+N_{\rm L}$), respectively.} \label{fig4_f_q_M}
\end{figure*}

\begin{figure*}
  \centering
  % Requires \usepackage{graphicx}
  \includegraphics[width=18cm]{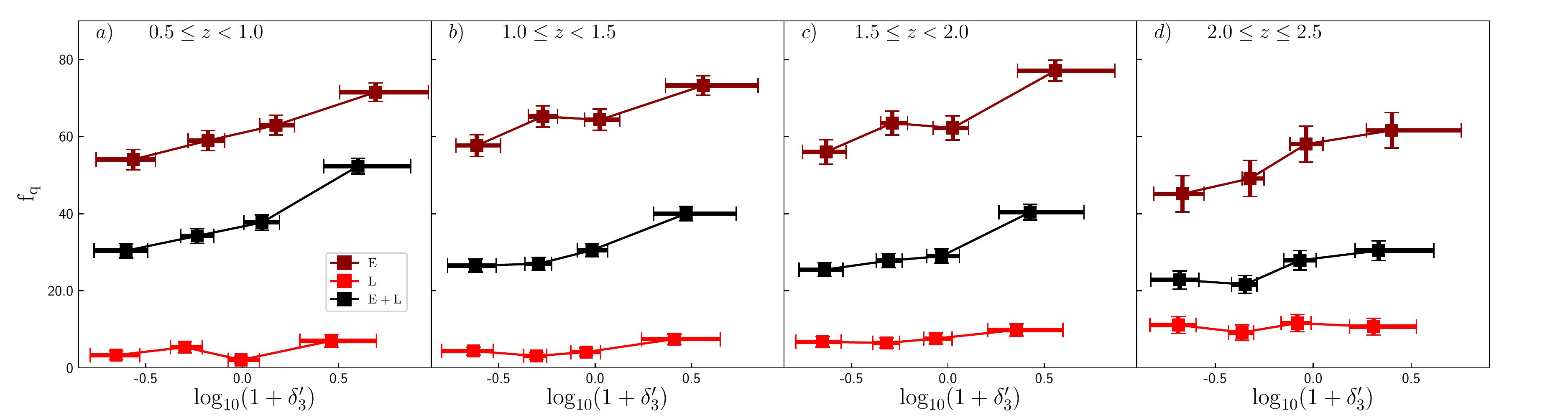}\\
  \caption{Quiescent factions as a function of local overdensity in four redshift intervals. Similar to Figure \ref{fig4_f_q_M}, the x-axis error bars indicate the range of 25th to 75th percentiles at each mass bin. The error bars of $f_{\rm q}$ indicate the uncertainty based on binomial distribution.} \label{fig5_f_q_env}
\end{figure*}

The quiescent fractions in different morphological types are employed as the indicator of star formation quenching proposed by \cite{Liu+2019}. In this section, to study the impact of stellar mass and environment on quenching process, the quiescent fractions out of ETGs ($f_{\rm q}(E) = \frac{N_{\rm qE}}{N_{\rm E}}$), LTGs ($f_{\rm q}(L) = \frac{N_{\rm qL}}{N_{\rm L}}$) and the total galaxies ($f_{\rm q}(E+L) = \frac{N_{\rm qE}+N_{\rm qL}}{N_{\rm E}+N_{\rm L}}$) are  denoted in darkred, red, and black colors as a function of stellar mass and local overdensity in different $z$-bins are shown in Figure \ref{fig4_f_q_M} and Figure \ref{fig5_f_q_env} respectively. Considering the binomial statistical distributions, the error of quiescent fraction can be computed as $\sigma_f = [f_{\rm q}(1 - f_{\rm q})/N_{\rm gal}]^{1/2}$, where $f_{\rm q}$ and $N_{\rm gal}$ are the quiescent fraction and the number of galaxies in subsamples, respectively.

In each panel of Figure \ref{fig4_f_q_M}, the total quiescent fraction colored by black lines, $f_{\rm q}(E+L)$, are first analyzed.  With the domination of QGs at the high-mass region, we find a strong mass dependence of star formation quenching.
With redshifts decrease, the fraction $f_{\rm q}(E+L)$ grows gradually, whereas the mass dependence is weakened due to a more significant increase of quiescent fraction at low-mass region.
The quenching of star formation seems to occur at high-mass region first, then shifts to less massive galaxies with cosmic time.
It confirms the  ``downsizing'' scenario on the evolution of galaxies \citep{Cowie+1996,Noeske+2007, Peng+2010, Brammer+2011, Goncalves+2012,Muzzin+2013, Gu+2018}.
%With the decreasing redshift, the fraction $f_q(E+L)$ increases gradually, whereas the mass dependence is weakened slowly due to the gradual growth of quiescent fraction at lower-mass region.

The quiescent fractions for ETGs ($f_{\rm q}(E)$, darkred lines) and LTGs ($f_{\rm q}(L)$, red lines) are also present in the Figure \ref{fig4_f_q_M}, respectively.
It is clear that $f_{\rm q}(E)$ is much higher than $f_{\rm q}(L)$, showing that the quenching process can be accompanied by the buildup of spheroidal component. Besides, the mass dependence of $f_{\rm q}(E)$ and $f_{\rm q}(L)$ exists as well, while it becomes less obvious with decreasing redshifts, especially for $z<1.0$.
Therefore, it may indicate that the in-situ process (stellar mass) takes a greater responsibility for the cessation of star formation at high redshifts, while it becomes less significant over cosmic time.

\begin{figure*}
  \centering
  % Requires \usepackage{graphicx}
  \includegraphics[width=18cm]{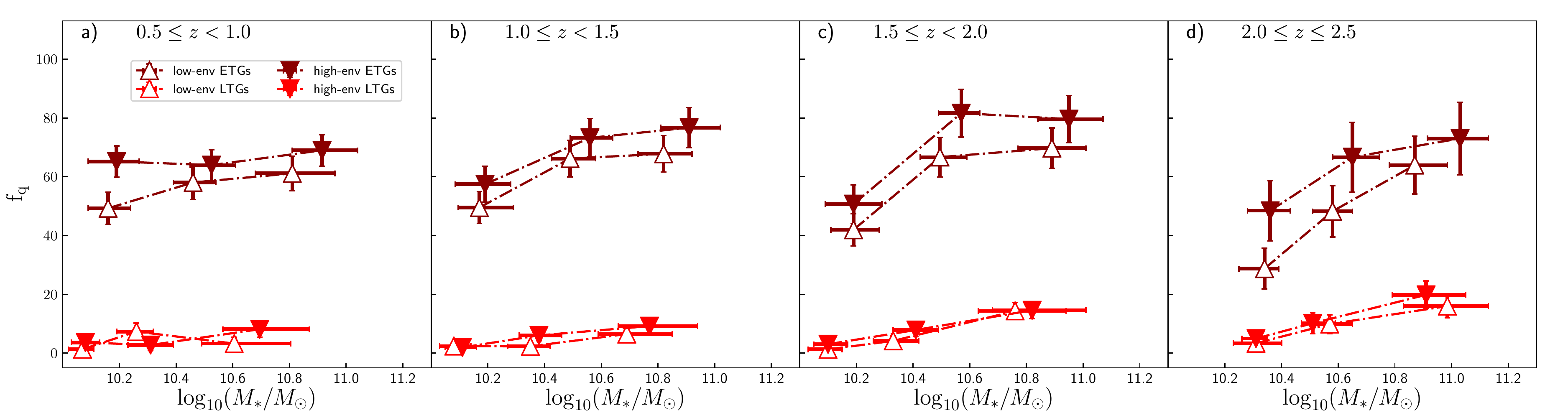}\\
  \caption{The quiescent fractions as a function of stellar mass at fixed local overdensity. The low- and high-environment are divided by the medians value of local overdensity for ETGs and LTGs at $0.5 \leq z \leq 2.5$. %ETGs and LTGs are split into two different local overdensity bins with roughly equal numbers of galaxies. 
  ETGs (LTGs) in high- or low-environment bins are denoted by solid or hollow symbols in darkred (red) color. The x-axis errors show the 25th to 75th percentiles. The error bars of $f_{\rm q}$ indicate the uncertainty based on binomial distributions. }\label{fig6_fq_M_env}
\end{figure*}

\begin{figure*}
  \centering
  % Requires \usepackage{graphicx}
  \includegraphics[width=18cm]{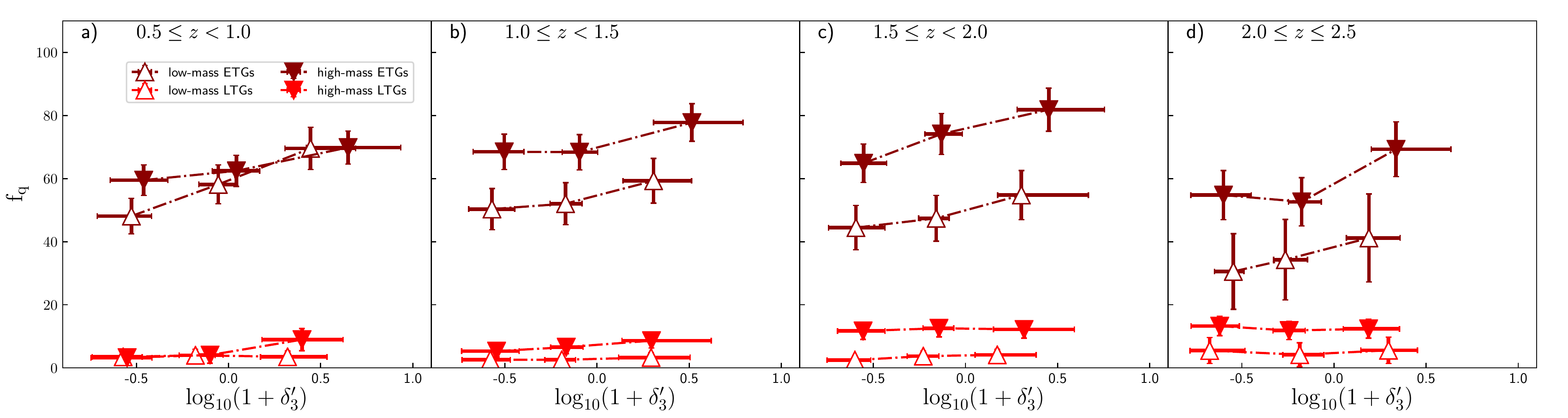}\\
  \caption{The quiescent fractions as a function of local overdensity at fixed stellar mass. The median values of stellar mass for ETGs and LTGs at $0.5 \leq z \leq 2.5$ are used to define the galaxies in low- and high-mass bins. %ETGs and LTGs are split into two different stellar mass bins with roughly equal numbers of galaxies. 
  ETGs (LTGs) in high- or low-mass bins are denoted by solid or hollow symbols in darkred (red) color. The x-axis errors show the 25th to 75th percentiles. The error bars of $f_{\rm q}$ indicate the uncertainty based on binomial distributions. }\label{fig7_fq_env_M}
\end{figure*}

The quiescent fractions as a function of local overdensity are shown in Figure \ref{fig5_f_q_env}. In a given redshift range, the QGs are more inclined to inhabit in a denser environment (i.e., $\log(1+\delta_3^\prime) > 0$). The highest environmental density is three times denser than its standard level (i.e., $\log(1+\delta_{3}^{\prime}) \sim 0.5$), which is the typical overdensity of clusters. For $f_{\rm q}(E+L)$, it shows a environmental dependence. Over cosmic time, it becomes stronger, especially at $z \sim 0.5$, which is supported by previous works, suggesting that the suppression of star formation can be regulated by environment-related events, and that such events come to play a dominate role since $z \sim 0.5$ (e.g., \citealt{BO+1978,Pandya+2017,Gu+2018}).

The distribution with environmental density for $f_{\rm q}(E)$ is similar to that of $f_{\rm q}(E+L)$, where environment gradually plays a more important role in star formation quenching towards lower redshifts. However, the environmental dependence for $f_{\rm q}(L)$ is less clear at the whole redshift ranges, which might attribute to the smaller number of qLs listed in Table \ref{table1:numbers of four types}.

In Figure \ref{fig4_f_q_M} and \ref{fig5_f_q_env}, we find that QGs tend to locate at higher stellar mass and local overdensity range. Since galaxies with higher stellar mass inclined to yield to denser environment, this result might be caused by the degeneracy between stellar mass and local overdensity. To check this problem, we re-calculate the quiescent fractions in ETGs and LTGs as a function of stellar mass at fixed local overdensity bins in Figure \ref{fig6_fq_M_env}. The low and high environment are divided by the median values of local environment distributions for ETGs and LTGs at $0.5 \leq z \leq 2.5$. The quiescent fraction as a function of local overdensity at fixed stellar mass bins is shown in Figure \ref{fig7_fq_env_M}. The median values of stellar mass at $0.5 \leq z \leq 2.5$ are used to define the low and high mass bins for ETGs and LTGs.

To check this, we re-calculate the quiescent fractions in ETGs and LTGs after spliting into two smaller bins of local overdensity and stellar mass in Figure \ref{fig6_fq_M_env} and \ref{fig7_fq_env_M}.

%we calculate the quiescent fractions as functions of stellar mass at fixed local environment bins in Figure \ref{fig4a_fq_M_env}, and the quiescent fraction at various local densities at fixed mass bins in Figure \ref{fig4a_fq_env_M}.}}

As shown in Figure \ref{fig6_fq_M_env}, a strong mass dependence of $f_{\rm q}$ still exists when considered in a fixed local overdensity bin. Along with cosmic time, mass dependence becomes less obvious, which in good agreements with previous works from local universe out to $z \sim 3$ \citep{Baldry+2006, van_den_Bosch+2008a, van_den_Bosch+2008b,Peng+2010, Balogh+2016, Grutzbauch+2011b, Grutzbauch+2011a, Kawinwanichakij+2017}.

\begin{figure*}
  \centering
  % Requires \usepackage{graphicx}
  \includegraphics[width=18cm]{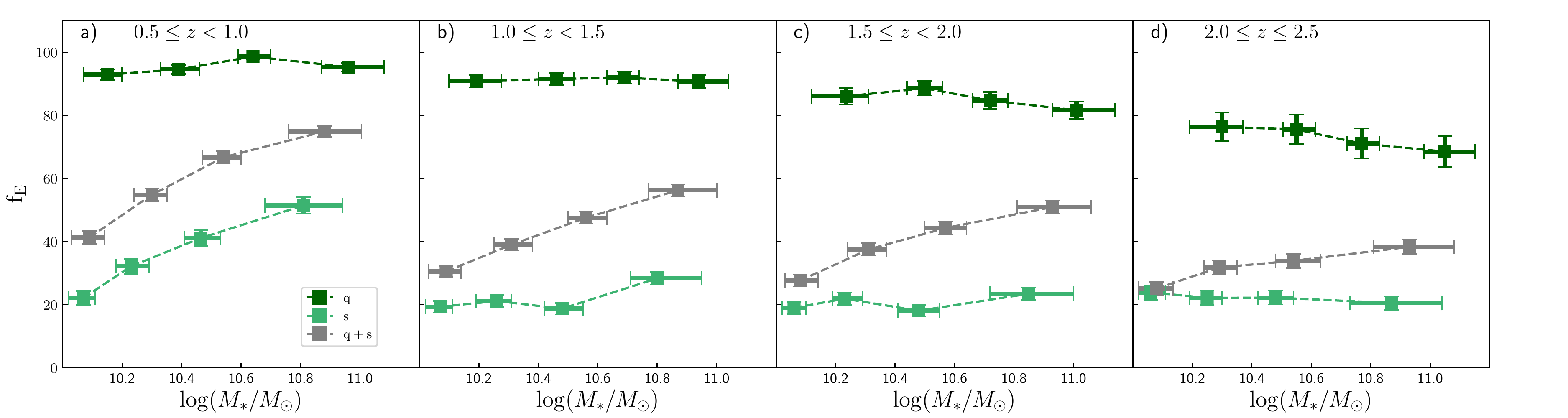}\\
  \caption{Early-type fractions out of QGs ($f_{\rm E}(\rm q) = \frac{N_{\rm qE}}{N_{\rm q}}$), SFGs ($f_{\rm E}(\rm s) = \frac{N_{\rm sE}}{N_{\rm s}}$) and both two populations ($f_{\rm E}(\rm q+s) = \frac{N_{\rm qE}+N_{\rm sE}}{N_{\rm q}+N_{\rm s}}$) at different redshift intervals. The $f_{\rm q}$, $f_{\rm s}$ and $f_{\rm (q+s)}$ in each panel are colored in green, lightgreen and grey separately. Split into four stellar mass bins with the similar galaxy numbers, we can derive the 25th and 75th percentile as errors of x-axis in each stellar mass bin. Assuming binomial distribution, the statistic uncertainly of each fraction is $\sigma_f = [f(1 - f)/N_{\rm tot}]^{1/2}$, where $f$ is representative to $f_{\rm E}(\rm q)$, $f_{\rm E}(\rm s)$ and $f_{\rm E}(\rm q+s)$ while $N_{\rm tot}$ is $N_{\rm q}$, $N_{\rm s}$ and $(N_{\rm q}+N_{\rm s})$ respectively.}\label{fig8_f_E_M}
\end{figure*}

\begin{figure*}
  \centering
  % Requires \usepackage{graphicx}
  \includegraphics[width=18cm]{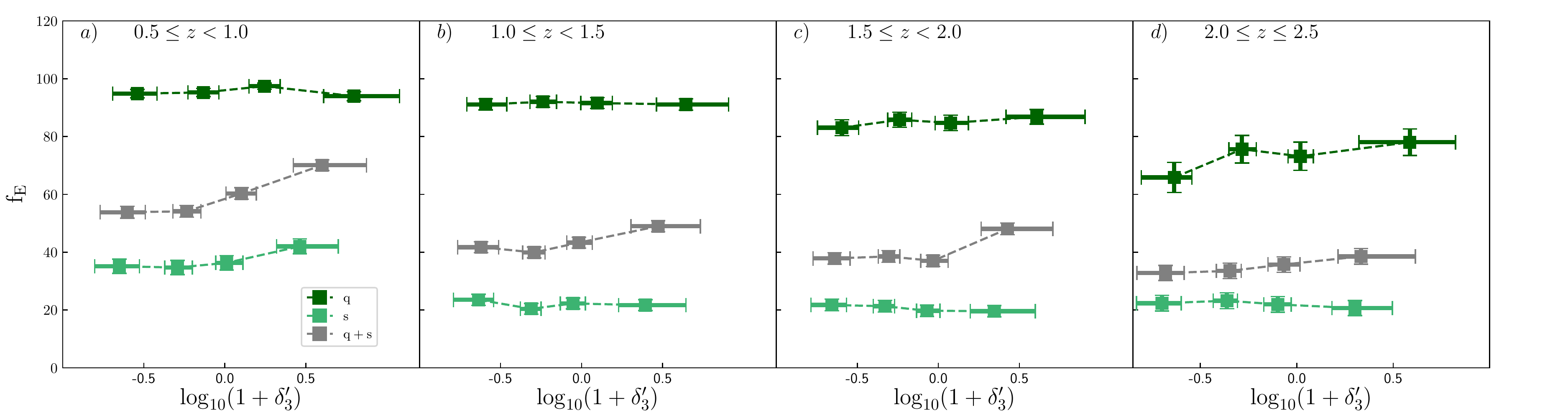}\\
  \caption{Early-type fractions as a function of local overdensity with cosmic time. Same as Figure \ref{fig8_f_E_M}, the x-axis errors indicate the 25th to 75th percentile. The error bars of $f_{\rm E}$ indicate the uncertainty based on binomial distribution.} \label{fig9_f_E_env}
\end{figure*}

In Figure \ref{fig7_fq_env_M}, at fixed mass bins, the QGs are also inclined to locate at denser environment. This proves that the environment dependence of $f_{\rm q}$ still exists, which is not affected by the mass distribution difference between SFGs and QGs, especially at $z<1.0$. \cite{Kawinwanichakij+2017} also showed that QGs are more common in overdense regions compared to SFGs even taking into account the differences in redshift and stellar mass. However, although a similar star formation-density relation is confirmed at $z<1.2$ in many papers \citep{Baldry+2006,van_den_Bosch+2008b, Peng+2010}, no environment dependence of 
$f_{\rm q}$ is found by \cite{Grutzbauch+2011b, Grutzbauch+2011a}, which may be caused by different environment density tracer, redshift range and selection criteria of QGs.

In general, the dependence of star formation quenching on stellar mass and environmental density can be found. These results point to an evolutionary trend that the quenching process of star formation is mainly regulated by stellar mass at high redshifts, while the environmental condition begins to dominate this process toward lower redshifts.

\section{Morphological transformation}\label{sec5:morphology transformation}

To figure out the influences of stellar mass  and environmental overdensity on morphological transformation, we compute the early-type fractions out of QGs ($f_{\rm E}(\rm q) = \frac{N_{\rm qE}}{N_{\rm q}}$), SFGs ($f_{\rm E}(\rm s) = \frac{N_{\rm sE}}{N_{\rm s}}$) and both two populations ($f_{\rm E}(\rm q+s) = \frac{N_{\rm qE}+N_{\rm sE}}{N_{\rm q}+N_{\rm s}}$) colord in green, lightgreen and grey in Figure \ref{fig8_f_E_M} and Figure \ref{fig9_f_E_env}. The error of early-type fraction is considered as the binomial statistical distributions.

In Figure \ref{fig8_f_E_M}, the $f_{\rm E}(\rm q)$ is overall much higher than $f_{\rm E}(\rm s)$, which proves the connection between morphological transformation and star formation quenching again. For the total early-type fractions (i.e., $f_{\rm E}(\rm q+s)$), a mass dependence of morphological transformation is identified in the given $z$-bins, and it becomes more evident over cosmic time. This implies that the formation of spheroidal component is previously completed at high-mass end particularly at low redshifts.

\begin{figure*}
  \centering
  % Requires \usepackage{graphicx}
  \includegraphics[width=18cm]{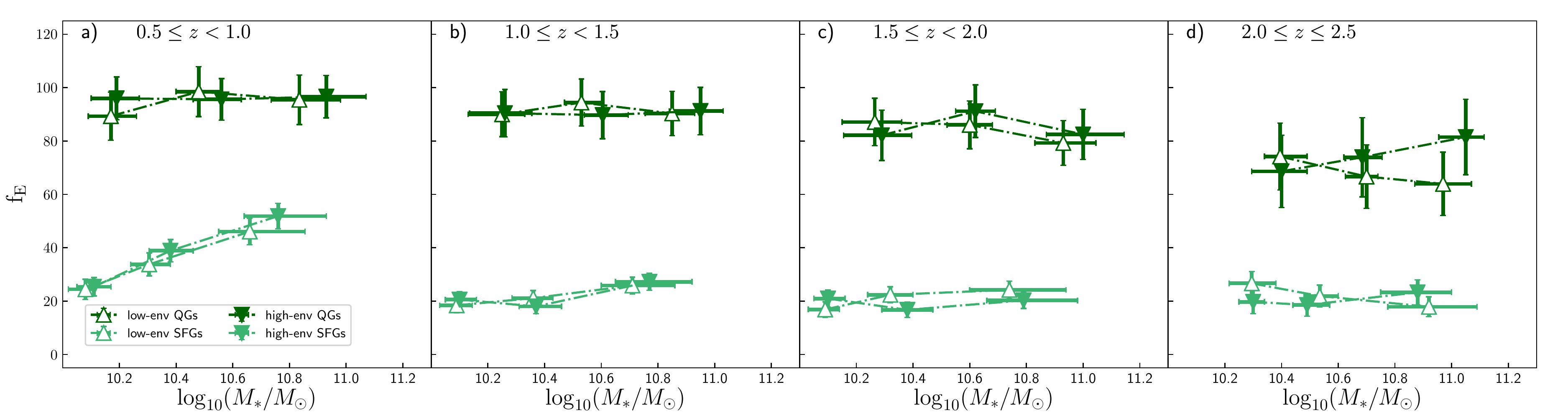}\\
  \caption{The early-type fractions as a function of stellar mass at fixed local overdensity and redshift bins. The QGs and SFGs are divided into low- and high-environment bins by the median values of local overdensity at $0.5 \leq z \leq 2.5$. QGs (SFGs) in high- or low-environment bins are denoted by solid or hollow symbols in green (lightgreen) color. The x-axis errors show the 25th to 75th percentiles. The error bars of $f_{\rm E}$ indicate the uncertainty based on binomial distributions.} \label{fig10_fE_M_env}
\end{figure*}

\begin{figure*}
  \centering
  % Requires \usepackage{graphicx}
  \includegraphics[width=18cm]{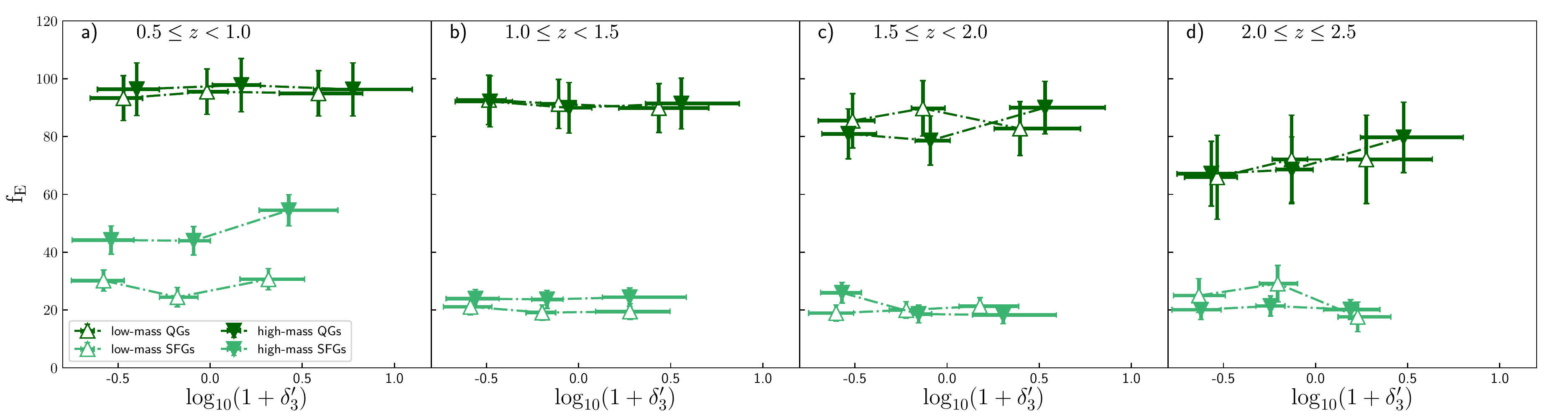}\\
  \caption{The early-type fractions as a function of local overdensity at fixed stellar mass. The low- and high-mass bins are defined by the median values of stellar mass for QGs and SFGs at $0.5 \leq z \leq 2.5$. QGs (SFGs) in high- or low-mass bins are denoted by solid and hollow symbols in green (lightgreen) color. The x-axis errors show the 25th to 75th percentiles. The error bars of $f_{\rm E}$ indicate the uncertainty based on binomial distributions.} \label{fig11_fE_env_M}
\end{figure*}

In the star-forming population, massive galaxies have larger early-type fractions especially for $z<1$, which demonstrates that the morphological transformation completes more frequently in massive galaxies when they are actively star-forming. However, this mass dependence of morphological transformation is unclear in QGs. Using the IllustrisTNG simulation, \cite{Tacchella+2019} proposed that galaxies have roughly built up the spheroidal component when their star formation being suppressed. Moreover, massive galaxies grow their spheroidal component more rapidly in star-forming phase, and the mature morphologies are formed once their star formation stops (see Figure 9 in \citealt{Tacchella+2019}). This picture coincides with our findings that a significant mass dependence exists in star-forming population, while it is not remarkable in QGs. %And, we will discuss the timescales of morphological transformation and star formation quenching in Section \ref{sec7:discussion}.

For early-type fractions at various local overdensities in Figure \ref{fig9_f_E_env}, the overall early-type fractions (i.e., $f_{\rm E}(\rm q+s)$) exist an obvious environment dependence since $z < 2$, reinforcing the morphology-density relation that the early-type fractions increases with environment densities in the local universe \citep{Norberg+2002,Goto+2003,Wolf+2007,Ball+2008,Joshi+2020}. The highest early-type fraction is located at the densest environment at $z<1$. Ram-pressure stripping of inter-cluster medium at denser environment can strip cold gas on a short timescales, and truncate the star formation of galaxies \citep{Quilis+2000}, which makes the spheroidal component more prominent.

However, the relations between early-type fractions in QGs and SFGs (i.e., $f_{\rm E}(\rm q)$ and $f_{\rm E}(\rm s)$) and local density are less clear. No statistical environment difference of $f_{\rm E}$ indicates that the morphological transformation is not affected by environmental condition at $0.5<z<2.5$.

To disentangle the influences of stellar mass and environment, we consider the early-type fractions in QGs and SFGs as a function of stellar mass at fixed local overdensity bins in Figure \ref{fig10_fE_M_env}, and as a function of local environment at fixed mass bins in Figure \ref{fig11_fE_env_M}. The low and high environment bins are divided by the median values of local overdensity at $0.5 \leq z \leq 2.5$ for QGs and SFGs. Similarly, the median values of stellar mass at $0.5 \leq z \leq 2.5$ are used to define the low and high mass bins for QGs and SFGs.

In Figure \ref{fig10_fE_M_env}, the distributional difference of $f_{\rm E}$ between high- and low-environment is negligible. With the domination of early-type population in massive regions, we confirm the strong mass dependence of $f_{\rm E}(\rm s)$ for $0.5<z<1.0$. As analyzed in Figure \ref{fig8_f_E_M}, no environment dependence of $f_{\rm E}(\rm q)$ is found, which might due to the fact that morphological transformation happens earlier than star formation quenching process proposed by \cite{Tacchella+2019}. Similar results are verified by \cite{Kawinwanichakij+2017} and \cite{Ownsworth+2016}, who used the S\'{e}rsic index $n$ as the morphological tracer. In Figure \ref{fig11_fE_env_M}, no clear environment dependence of $f_{\rm E}$ is shown at a fixed stellar mass. Based on different environment tracer and classification of QGs, morphological transformation seems to have not obviously influenced by environment condition \citep{van_den_Bosch+2008b, Grutzbauch+2011b, Grutzbauch+2011a, Kawinwanichakij+2017}.

In summary, the $f_{\rm E}$ behaviors in Figures \ref{fig8_f_E_M}, \ref{fig9_f_E_env}, \ref{fig10_fE_M_env}, \ref{fig11_fE_env_M} suggest that morphological transformation is mainly regulated by stellar mass at low redshifts, and it seems insensitive to the surrounding environmental densities. Extended to the local universe, our results should still hold according to the morphological analysis in \cite{Liu+2019}.

\section{AGN fractions}\label{sec6:AGN fractions}

The GOODS-N and GOODS-S fields are covered by the 2 Ms Chandra Deep Field-North (CDF-N), the 7 Ms Chandra Deep Field-South (CDF-S) surveys. Based on these surveys, \cite{Xue+2016} and \cite{Luo+2017} provided their estimations in three standard X-ray bands: 0.5-7 keV (full band), 0.5-2.0 keV (soft band) and 2.0-7.0 keV (hard band), and performed an identification of galaxy type (``AGN",``GALAXY", or ``STAR").  AGN can be identified by one of the following six criteria: (1) $L_{\rm X,int} \geq 3 \times 10^{42} ~ \rm erg ~ s^{-1}$, where $L_{\rm X,int}$ is the absoption-corrected intrinsic luminosity in full X-ray band; (2) $\Gamma \leq 1.0$, where $\Gamma$ is the effective power law index of photon, for selecting obscured AGNs; (3) $\log(f_{\rm X}/f_{\rm R}) > -1$, where $f_{\rm X}$ is the flux for soft, hard or full band, and $f_{\rm R}$ is the R-band flux; (4) $L_{\rm X,int}/L_{\rm 1.4GHz} \geq 2.4 \times 10^{18}$; (5) spectroscopically classified as AGNs; (6) $\log(f_{\rm X}/f_{\rm Ks}) > -1.2$, which is only used for the CDF-S field with the Ks-band photometry.

\begin{figure}
  \centering
  % Requires \usepackage{graphicx}
  \includegraphics[width=7cm]{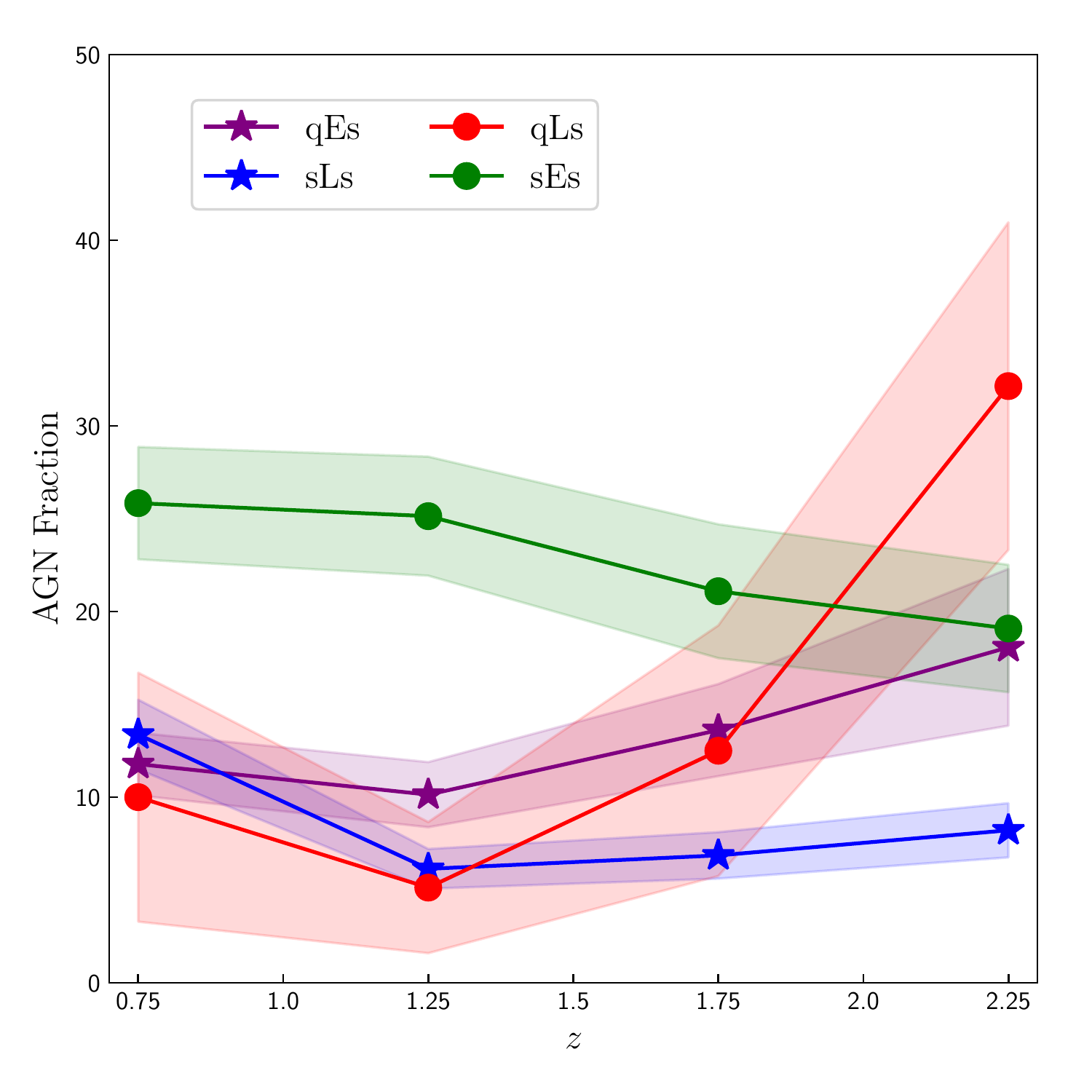}\\
  \caption{The AGN fractions in qEs, qLs, sEs and sLs subsamples as a function of redshift, denoted in red, yellow, blue and lightblue colors, respectively. The shaded region represents the uncertainty of $f_{\rm AGN}$ for each type of galaxies.}\label{fig12_agn}
\end{figure}

Within a searching radius of $1.5''$, we identify the host galaxies of AGNs in the GOODS-N and GOODS-S fields. To testify AGN effects on the formation of galaxies, we calculate the AGN fractions for our four galaxy types separately. The AGN fractions and their corresponding errors are shown in Figure \ref{fig12_agn}, as a function of redshift. Considering the binomial statistical distributions, the error bars are computed as $\sigma_f = [f_{\rm AGN}(1 - f_{\rm AGN})/N_{\rm tot}]^{1/2}$, where the $f_{\rm AGN}$ and $N_{\rm tot}$ are the AGN fraction and total number of galaxies in each redshift bin for four subsamples.

Apparently, quiescent late-type galaxies (i.e., qLs), where star formations are suppressed but spheroidal components are not predominated, exhibit the highest AGN fractions at $z>2$, and then drop rapidly until $z \sim 1.25$. Similarly, it is found that the higher AGN fractions also exhibit in red sequence and green valley at early epoch \citep{Nandra+2007,Salim+2007,Schawinski+2010,Gu+2018}, whose star formations are entirely or partially truncated. AGN activity is thought to drive out or heat cold gas to suppress star formations and maintain the quiescence of galaxies before the morphological transformation at $z \sim 2$  \citep{Wang+2017,Gu+2018}. In this scenario, negative AGN feedback could play a role in the formation of qLs. Over cosmic time, the decrement of AGN fractions in qLs could be attributed to the reduced AGN activities.

In addition, at $z < 2$, the highest AGN fraction is found in the sEs, and it increases slightly with cosmic time. These galaxies have built up their spheroidal components, but their star formations are still prevalent.   This can be interpreted by the picture that the morphological transformation in sEs is associated with some previous gas-rich violent events (e.g., merger or disk instability) that subsequently triggered starbursts \citep{Toomre+1977,Hopkins+2008}. 
%Similar to sEs, compact SFGs(cSFGs) are chosen by their higher star-forming activities and compactness morphologies, such as $\Sigma_{1.5} = M_*/\rm r_e^{1.5} > 10.3 ~ M_{\odot}\rm kpc^{-1.5}$ defined by \cite{Barro+2013}. As is known, cSFGs are the progenitors of compact QGs(cQGs), and they possess the highest AGN fractions of $\sim 30\%$ \citep{Barro+2013,Gu+2019,Gu+2020,Lu+2019,Lu+2020}. The formation of cSFGs is usually of cSFGs is usually associated with some previous gas-rich violent events(e.g., merger or disk instability) that subsequently triggered an AGN \citep{Fang+2015}.
In this scenario, AGN plays a positive feedback to star formation.
Hence, on account of the highest AGN fractions in this work and the larger asymmetry indices noted by \cite{Liu+2019} in sEs, we are also in favor of the positive AGN feedback, and suggest that the sEs are likely to be the remnants of gas-rich mergers. As for sLs and qEs, their AGN fractions are lower than 15\%, showing no prominent connection between their formations and AGN activities.

\begin{figure}
  \centering
  % Requires \usepackage{graphicx}
  \includegraphics[width=7cm]{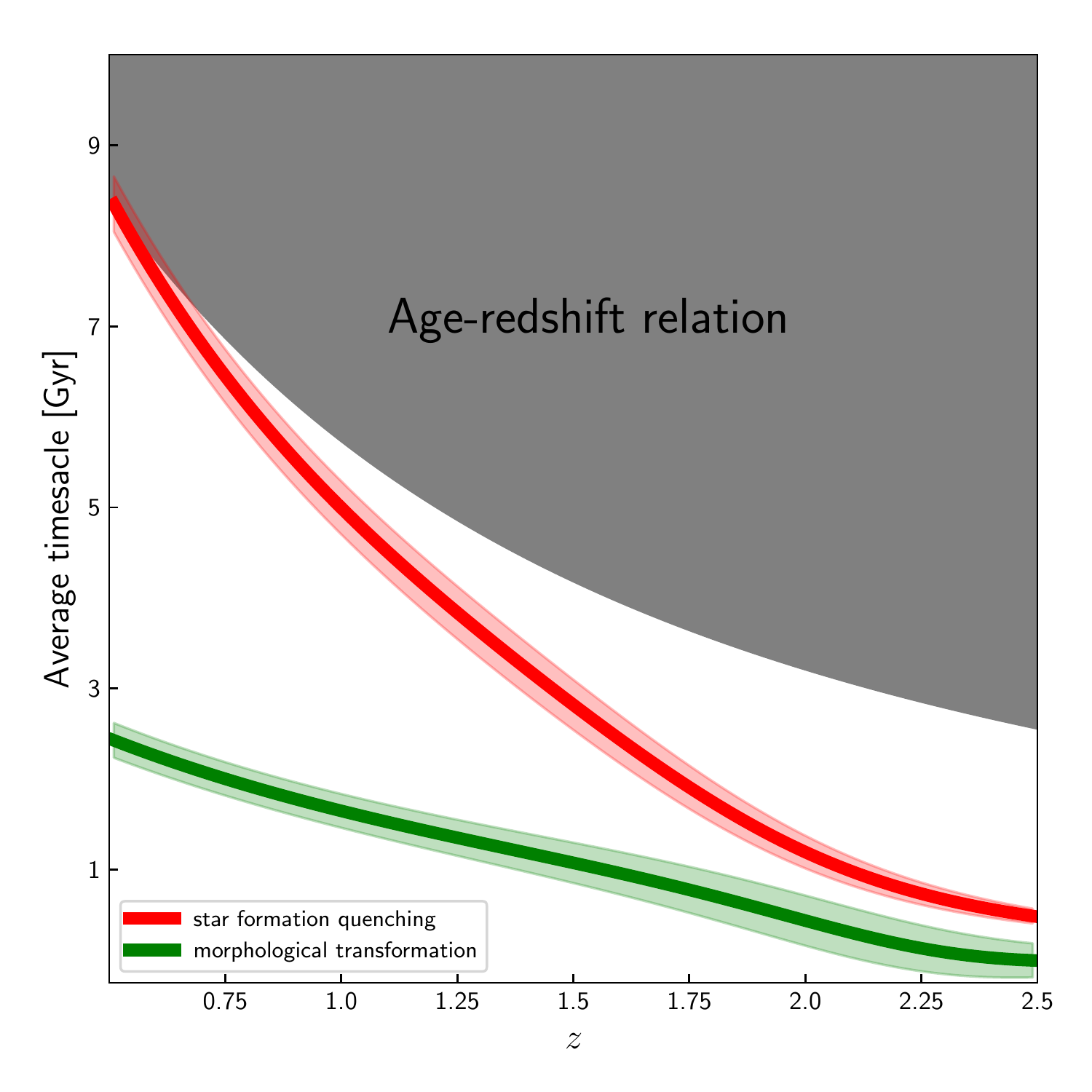}\\
  \caption{Redshift evolution of morphological transformation(green) and star formation quenching(red). Green and red shaded regions are the propagated errors. The grey shaded region indicates the timescales that above the age-redshifts relation.}\label{fig13_timescale}
\end{figure}

\section{Discussion}\label{sec7:discussion}
\subsection{\rm Redshift Evolution of Transition Timescales} \label{sec7.1:timescales}
To quantify the difference in timescales between star formation quenching and morphological transformation, we estimate the average transition timescales following \cite{Pandya+2017}.

By employing a cubic polynomial fits to observed number densities of blue, green and red populations as a function of redshifts, the average transition timescale can be derived at given redshifts.
The average quenching timescale between $z_1$ and $z_2$ ($\langle t_{\rm transition} \rangle _{z_1,z_2}$) is evaluated as 

\begin{equation}\label{func5_timescale}
  \langle t_{\rm transition} \rangle _{z_1,z_2} = \langle n_{\rm transition}\rangle _{z_1,z_2} \times (\frac{dn_{\rm quiescent}}{dt})_{z_1,z_2}^{-1},
\end{equation}
where $\langle n_{\rm transition}\rangle _{z_1,z_2}$  is the average number density of transition population (i.e., green valley galaxies) and $(\frac{dn_{\rm quiescent}}{dt})_{z_1,z_2}^{-1}$ is the change of number density in QGs with regard to the cosmic time from $z_1$ to $z_2$. However, it should be noted that the average quenching timescale gives an upper limit of quenching timescale because this method is based on an extreme assumption that all green valley galaxies evolve from star-forming, and transform to quiescent population monodirectionally. Rejuvenation events or  oscillatory excursions from star formation main sequence have not been taken into consideration.

The massive ($\log M_*/M_{\odot} \geq 10$) galaxies can be classified into blue, green and red populations according to the dust-corrected colors out to $z = 2.5$ (see \citealt{Wang+2017,Gu+2018} for  detail). We have calculated the average quenching timescale as a function of redshift \citep{Gu+2019}, which is denoted by the red line in Figure \ref{fig13_timescale}. The result explicitly indicates that the quenching process happens in a shorter timescale at high redshifts, and the timescale  becomes longer with decreasing redshifts, which is in a good agreement with previous works \citep{Barro+2013,Pandya+2017}.

Inspired by the method of deriving  the average quenching timescale, we also evaluate the average timescale of morphological transformation. The great importance is  to select a transition population in morphological transformation, just like the green valley population during star formation quenching. As mentioned in Section \ref{sec2.2:Early type vs. Late type}, \cite{Huertas-Company+2015} have proposed  five morphological classifications (namely, [SPH], [DSPH], [DISK], [DIRR], and [IRR]) with total numbers of 2358, 1258, 1768, 2630, and 930, respectively. We adopt a  monodirectional model of morphological transformation, assuming that irregular-dominated galaxies (i.e., [IRR] and [DIRR]) would evolve to disk-dominated population (i.e., [DISK]), and transform into spheroidal-dominated populations (i.e., [DSPH] and [SPH]). We do not take the intense merger events or regrowth of disks into account. Similarly, by adopting the disk-dominated galaxies as the transition population, we can estimate the average timescale of morphological transformation by

\begin{equation}\label{func6_mor_timescale}
  \langle t_{\rm transition} \rangle _{z_1,z_2} = \langle n_{\rm [DISK]}\rangle _{z_1,z_2} \times (\frac{dn_{\rm [DSPH]+[SPH]}}{dt})_{z_1,z_2}^{-1}.
\end{equation}
In Figure \ref{fig13_timescale}, the redshift evolution of morphological transformation timescales is shown by the green line. The morphological transition timescale increases slightly with cosmic time. But basically, the morphological transformation spends a shorter time than the quenching process. It is conspicuous that the difference in timescales between morphological transformation and star formation quenching becomes larger over cosmic time.

\begin{figure}
  \centering
  % Requires \usepackage{graphicx}
  \includegraphics[width=7cm]{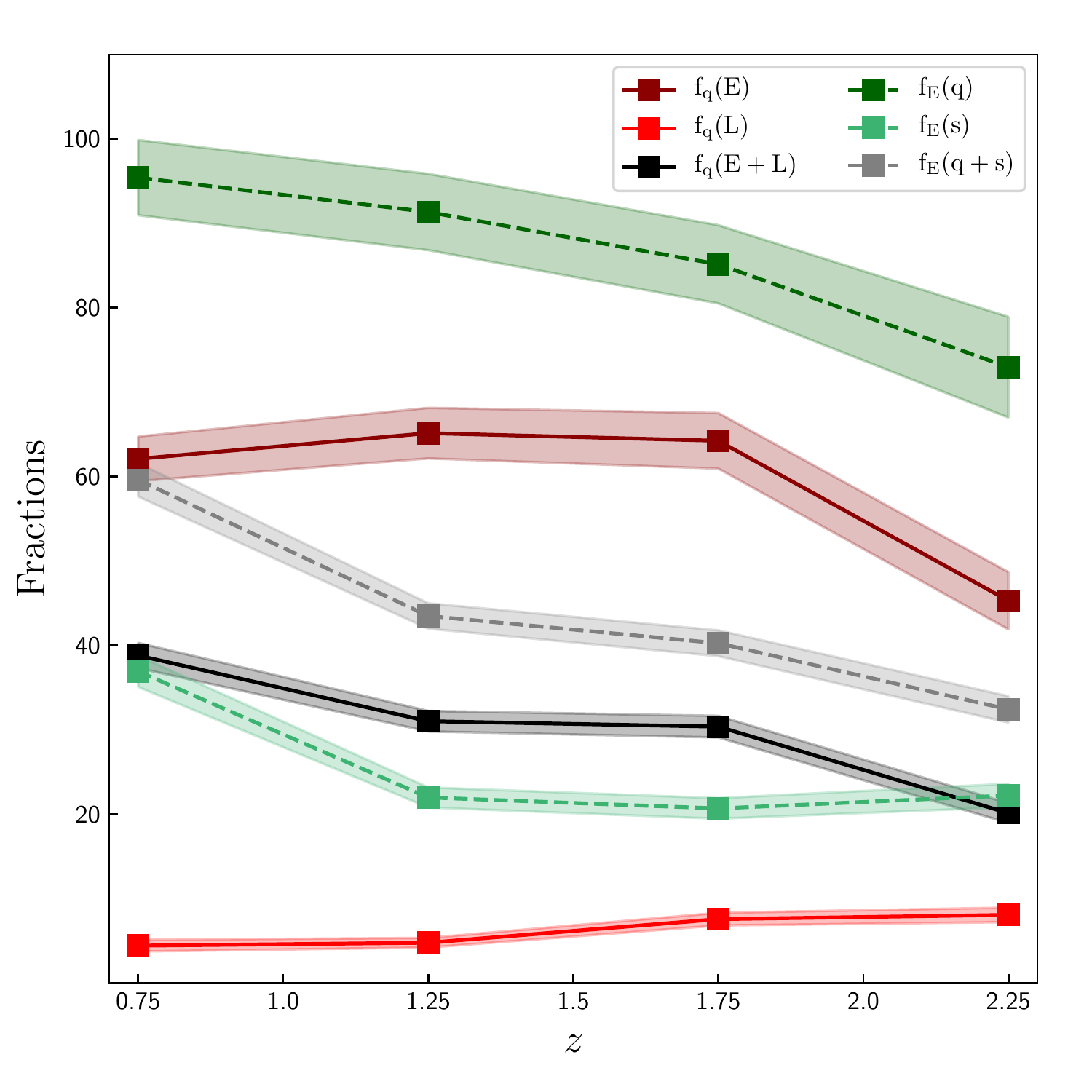}\\
  \caption{Quenched fractions out of early, late and total types denoted in darkred, red and black dots. Early type fractions out of quiescent, star forming and both two populations are shown in green, lightgreen and grey colors. Their errorbars are identical Figure \ref{fig4_f_q_M} and Figure \ref{fig8_f_E_M}.}\label{fig14_frac_z}
\end{figure}

To verify the reliability of our timescales, we further consider the redshift evolution of the $f_{\rm q}$ and $f_{\rm E}$  in Figure \ref{fig14_frac_z}.
The early-type fractions (shown in dot-dashed lines) increases with cosmic time, while the quiescent fractions (shown in solid lines) does not increase with cosmic time at a given morphological type. The total early-type fraction (colored in grey) is overall higher than total quiescent fractions (colored in black), and their difference is higher at later epoch.
This result supports that the shorter timescales in morphological transformation in lower redshifts  which is in consistent with the result in Figure \ref{fig13_timescale}. The shorter timescale of morphological transformation may lead to an increasing number of the sEs at lower redshifts.  Table \ref{table1:numbers of four types} shows that the number of sEs is greater than that of qLs. And we do find the a steeper slope of $f_{\rm E}(\rm s)$  at $z\sim 1$, which is probably due to the larger difference in timescales. It suggests that the morphological transformation might be accomplished much earlier than the suppression of star formation at low redshifts. %Our transition timescales approximately provide an reasonable explanation of the formation of sEs.

\subsection{\rm The Formation of qLs and sEs}\label{sec7.2:The formation of qLs and sEs}

``Merger hypothesis" is commonly connected with the explosive quasar or starburst phase, which leads to the morphological transition from rotation-dominated disks into pressure-dominated spheroids \citep{Toomre+1977,Hopkins+2008}. Strong star formation and black hole accretion are both triggered by the inflow of cold gas from the coalescence of gas-rich galaxies \citep{Springel+2005,Kaviraj+2009,George+2017}, which is considered as the so-called positive AGN feedback. However, once the static massive hot halo has formed, the accretion of shock heating low-energy gas towards the central supermassive black hole could result in the negative AGN feedback, which prevents gas from further cooling \citep{Croton+2006,Baldry+2008}. Meanwhile, interaction between galaxy and intercluster medium in a denser environment \citep{Moran+2007}, galaxy-galaxy harassment \citep{Moore+1999}, and starvation \citep{Larson+1980}, can be responsible for halting the supply of cold gas, ceasing the star formation, and rapidly building up the spheroidal component.

Star-forming early-type galaxies (sEs), that appear spheroid-dominated and undergo star formation, possess the highest AGN fractions especially at $z<2$. Since the difference in the environmental conditions between sEs and sLs is insignificant at $0.5 \leq z \leq 2.5$, we suppose that environment should have not played an important role on the formation of sEs. As explained in the scenario of positive AGN feedback, the galaxies experienced the violent gas-rich merger event would change their morphologies and trigger the activity of star formation. In this scenario, the morphological transformation is closely connected with the trigger of AGNs. According to Figure \ref{fig12_agn} and Figure \ref{fig13_timescale}, with respective to the quenching timescale, the morphological transition tends to be completed more efficiently at lower redshifts, which points to a higher AGN fraction in the sEs. An alternative origin of sEs is the rejuvenation of qEs,  and external acquisition of gas (e.g., through minor gas-rich mergers or intergalactic medium accretion) could be able to drive star formation again \citep{Kannappan+2009,Kim+2018,Liu+2019}. 
%Coincidently, it is found that the relative small variation with redshifts is presented in both the timescale of morphological transformation and the AGN fractions in sEs.  
Hence, we suggest that some sEs might be the remnants of gas-rich mergers driven by positive AGN feedback, or the results of rejuvenated qEs at $z<2$.

As illustrated in Figure \ref{fig12_agn}, the quiescent late-type galaxies (qLs) have the highest AGN fractions at $z>2$. It implies that negative AGN feedback could be effective at truncating the star formation and maintaining the quiescence of galaxies at $z>2$. It supports that negative AGN feedback might dominate the formations of qLs before the buildup of spheroidal component at early epoch. It is found that AGN fraction in qLs decreases along with cosmic time. Negative AGN feedback might be of less importance for the formation of qLs at low redshifts. In addition, unlikely the sLs at $z<1.5$, the qLs  prefer to locate at denser environment. It hints that dense environment might act in halting the cold gas supply at later epoch. In a word, the formation of qLs could be attribute to the negative AGN feedback at high redshifts, while it might be driven by environmental quenching since $z \sim 1.5$. %However, as shown in Figure \ref{fig1_UVJ}, the percentage of qLs out of the whole sample decreases towards lower redshifts.  shows that {\color{red} environmental effects may not competent to produce the qLs.  The former scenario of AGN negative feedback performs better in the formation of qLs than the latter environmental effects.}

\section{Summary}\label{sec8:summary}

In order to study the properties of star formation quenching and morphological transformation, we choose massive galaxies ($\log M_*/M_{\odot} \geq 10$) in five fields of 3D-HST at $0.5 \leq z \leq 2.5$.  Our sample is divided into four galaxy types: quiescent early-type (qEs), quiescent late-type (qLs), star-forming early-type (sEs), and star-forming late-type (sLs) galaxies. SFGs and QGs are identified by the $UVJ$ diagnosis, whereas early- and late-type galaxies are identified by the possibility whether they possess a spheroid-dominated component on the basis on \cite{Huertas-Company+2015}.  We analyze the stellar mass and local environment distributions of these four populations, and explore the mass and local environment dependence of star-formation quenching and morphological transformation. Moreover, to figure out the formation of sEs and qLs, we discuss the AGN fractions in four subsamples, and estimate the average timescales of morphological transformation and star formation quenching. Our results are summarized as follows:

(1) At the given morphology type, QGs present entirely different mass distributions at $z>1$ compared with the star-forming counterparts.
At the given star forming status (star-forming or quiescent), the difference of mass distribution between early- and late-type galaxies enlarge with cosmic time. Therefore, stellar mass plays an important role in star formation quenching at high redshifts, while it is significant to morphological transformation at later epoch.

(2) At a fixed morphological type, different local environment distributions between QGs and SFGs are found at $z<1.5$. Nevertheless, at the given star forming status, early- and late-type galaxies present the similar local overdensity distributions at $0.5<z<2.5$. It indicates that the local overdensity is significant in star formation quenching at later epoch, while it seems to be irresponsible for morphological transformation.

(3) The larger quiescent fractions are found at high-mass and high-density ends. The process of star formation quenching exhibits a strong dependence on stellar mass at early epoch, and the mass dependence of quenching tends to decrease with cosmic time. In addition, a crucial local environment dependence on star formation quenching is verified  only at $z < 1.0$.

(4) Massive galaxies are likely to possess higher early-type fractions. Morphological transformation shows a clear dependence on stellar mass but not on local environment. And the mass dependence of morphology transformation becomes stronger over cosmic time.

(5) The highest AGN fraction at $z>2$ is exhibited in qLs, which indicates the negative AGN feedback could be responsible for the formation of qLs. At $z<2$, the sEs are identified to have the highest AGN fraction. It suggests that the positive AGN feedback might contribute to the formation of sEs.

% With regard to the formation of qLs and sEs, we only demonstrate that AGNs and environment might play an important role. However, the differences of their physical properties are needed to pay a more detailed attention in further study.

~\\
%\acknowledgments

This work is based on observations taken by the 3D-HST Treasury Program (GO 12177 and 12328) with the NASA/ESA HST, which is operated by the Association of Universities for Research in Astronomy, Inc., under NASA contract NAS526555.
This work is supported by the National Natural Science Foundation of China (nos. 11873032, 11673004, 11433005) and by the Research Fund for the Doctoral Program of Higher Education of China (no.20133207110006). G.W.F. acknowledges the support from Chinese Space Station Telescope (CSST) Project. G.Y.Z acknowledges the support from China Postdoctoral Science Foundation (2020M681281) and Shanghai Post-doctoral Excellence Program (2020218).

\bibliography{my_bib.bib}
\bibliographystyle{aasjournal}
\end{document}